\documentclass[journal]{IEEEtran}

\usepackage{amsmath}
\usepackage{amssymb}
\usepackage{algorithm}
\usepackage{algpseudocode}
\usepackage{graphicx} 
\usepackage{multirow}
\usepackage{hhline}
\usepackage{enumitem}

\begin{document}
\pagestyle{plain}

\title{Multi-Material Decomposition Using Spectral Diffusion Posterior Sampling}
\author{Xiao~Jiang,
        Grace~J.~Gang,
        and~J.~Webster~Stayman  \vspace*{-1.0cm}

\thanks{This work has been submitted to the IEEE for possible publication. Copyright may be transferred without notice, after which this version may no longer be accessible.}
\thanks{Xiao Jiang and J.Webster Stayman are with the Department of Biomedical Engineering, Johns Hopkins University, Baltimore, MD, 20205 USA. e-mail: xjiang43@jhu.edu, web.stayman@jhu.edu}% <-this % stops a space
\thanks{Grace~J.~Gang is with the Department of Radiology, University of Pennsylvania, Philadelphia, PA, 19104 USA. e-mail: grace.j.gang@pennmedicine.upenn.edu}}

% make the title area
\maketitle

\begin{abstract}
Many spectral CT applications require accurate material decomposition. Existing material decomposition algorithms are often susceptible to significant noise magnification or, in the case of one-step model-based approaches, hampered by slow convergence rates and large computational requirements. In this work, we proposed a novel framework - spectral diffusion posterior sampling (spectral DPS) - for one-step reconstruction and multi-material decomposition, which combines sophisticated prior information captured by one-time unsupervised learning and an arbitrary analytic physical system model. Spectral DPS is built upon a general DPS framework for nonlinear inverse problems. Several strategies developed in previous work, including jumpstart sampling, Jacobian approximation, and multi-step likelihood updates are applied facilitate stable and accurate decompositions. The effectiveness of spectral DPS was evaluated on a simulated dual-layer and a kV-switching spectral system as well as on a physical cone-beam CT (CBCT) test bench. In simulation studies, spectral DPS improved PSNR by $27.49\%$ to $71.93\%$  over baseline DPS and by $26.53\%$ to $57.30\%$ over MBMD, depending on the the region of interest. In physical phantom study, spectral DPS achieved a $<1\%$ error in estimating the mean density in a homogeneous region. Compared with baseline DPS, spectral DPS effectively avoided generating false structures in the homogeneous phantom and reduced the variability around edges. Both simulation and physical phantom studies demonstrated the superior performance of spectral DPS for stable and accurate material decomposition.
\end{abstract}


\section{Introduction}
\IEEEPARstart{S}{pectral} CT uses energy-dependent attenuation information to enhance the clinical value of CT with a wide range of applications including tissue characterization\cite{wang2014quantitative}, quantification of exogenous contrast agents\cite{symons2017photon}, electron density/stopping power estimation\cite{ohira2018estimation}, and virtual non-contrast imaging\cite{sauter2018dual}.

Many of these applications rely on accurate material decomposition, which reconstructs basis material density maps from the spectral projection data\cite{long2014multi}. Material decomposition is an ill-conditioned nonlinear inverse problem without an explicit solution. Existing material decomposition algorithms can be categorized into three main types: analytical decomposition\cite{jiang2021fast}, iterative/model-based decomposition\cite{tilley2019model}, and learning-based decomposition\cite{gong2020deep}. Analytical algorithms, which apply pre-calibrated linear or polynomial functions to approximate material density, permit fast computation; however, they are often hampered by unmodeled beam-hardening effects and a poor signal-to-noise ratio, necessitating additional pre/post-processing \cite{petrongolo2015noise}  to enhance image quality. Model-based algorithms formulate the material decomposition as an optimization problem. Benefiting from an accurate underlying physics model, model-based algorithms can effectively mitigate beam hardening artifacts and suppress image noise. Nonetheless, the optimization process typically requires many iterations to converge due to the ill-conditioned nature of the estimation problem. Moreover, while conventional regularization is helpful in reducing noise, it can often alter the image texture and introduce undesired edge blur. 

Machine learning algorithms have been extensively used in medical image formation including spectral CT\cite{gong2020deep,wu2021deep,abascal2021material,clark2018multi,zhu2022feasibility}. Such approaches leverage prior knowledge learned from large datasets and can generally surpass classic approaches in terms of imaging accuracy and reconstruction speed. However, many methods do not explicitly leverage known imaging physics to enhance consistency with measurement data. Such approaches raise concerns about the robustness of image formation as well as potential network hallucinations, etc. To address this issue, some methods integrate a physical model as part of network training to improve data consistency\cite{fang2021iterative,eguizabal2022deep}. However, typically these trained networks are tailored to specific system models, requiring network retraining for changes in protocol, technique, geometry, or other device-specific factors.

Diffusion Posterior Sampling (DPS)\cite{chung2022diffusion} has established a new framework to integrate a learned prior and an accurate physical model of measurements. Specifically, DPS starts with unsupervised training of a score-based generative model (SGM) to capture the target domain distribution, then applies a reverse process to estimate image parameters - alternating between SGM reverse sampling to drive the image towards the target distribution and model-based updates to enforce the data consistency with the measurements. As a result, the final output adheres to both the prior distribution and measurements. A major advantage of DPS is that its network training is not specific to any physical model, allowing for application across different imaging systems, protocols, etc. Recently, we have applied DPS to nonlinear CT reconstruction\cite{li2023diffusion,jiang2024strategies}, demonstrating its effectiveness for both low-exposure and sparse-view CT reconstruction. 

In this work, we extended our DPS CT reconstruction framework to Spectral DPS - targeting joint reconstruction and material decomposition in spectral CT to accommodate different spectral system settings using a single neural network trained through unsupervised learning. The performance of spectral DPS is demonstrated on simulated kV-switching\cite{cassetta2020fast} and dual-layer\cite{wang2021high} CT systems, as well as a physical bench CBCT system. Preliminary results of this work were presented in \cite{jiang2024ct}. This work has been expanded to include parameter optimization, physical experiments, and quantitative analysis. This paper is organized as: Section \ref{sec:physics} introduces the physics model of spectral CT system. Section \ref{sec:dps} elaborates on the spectral DPS framework. Technique details about the network training and experiment settings are included in Section \ref{sec:train} to \ref{sec:eval}. Results are summarized in Section \ref{sec:results} while conclusion and discussion in Section \ref{sec:dis}.

\section{Methodology}
\subsection{Physics Model of Spectral CT System}
\label{sec:physics}
Various spectral CT system designs with significant differences in system geometry\cite{tivnan2022design,stayman2020grating}, detector physics\cite{wang2021high,willemink2018photon}, and spectral sensitivity\cite{tivnan2020combining} have been investigated in both the literature and clinical applications\cite{kang2010dual,van2023photon}. This work utilizes the general spectral CT model proposed in \cite{tilley2019model}, with system parameters and notation defined in TABLE.\ref{tab:system}. The measurements of each spectral channel is assumed to follow a multivariate Gaussian distribution. Leveraging matrix notation, the measurement model is written in the following compact form \cite{wang2021high}:

\begin{equation}
\label{eq:phyics_model}
    \textbf{y} \sim \mathcal{N}(\overline{\textbf{y}}, \textbf{K}), \quad \overline{\textbf{y}} = \textbf{BS}\exp({-\textbf{QAx}}),
\end{equation}

where

\[
\textbf{y} = \begin{bmatrix}
        \textbf{y}_1 \\
        \vdots\\
        \textbf{y}_J \\
        \end{bmatrix}
        ,
\textbf{x} = \begin{bmatrix}
        \textbf{x}_1 \\
        \vdots\\
        \textbf{x}_K \\
        \end{bmatrix}
        ,
\textbf{K} = \begin{bmatrix}
        \textbf{K}_1&& \\
        &\ddots&\\
        &&\textbf{K}_J \\
        \end{bmatrix}
\]
\[
\textbf{B} = \begin{bmatrix}
        \textbf{B}_1\\
        & \ddots &\\
        &  & \textbf{B}_J\\
        \end{bmatrix}
,
\textbf{S} = \begin{bmatrix}
        \textbf{S}_1^T \otimes \textbf{I}^{P_1} \\
        & \ddots &\\
        &  & \textbf{S}_J^T \otimes \textbf{I}^{P_J} \\
        \end{bmatrix}
\]
\[
\textbf{Q} = \begin{bmatrix}
        \textbf{I}^J \otimes \textbf{Q}_1 \otimes \textbf{I}^{P_1},  & \hdots, & \textbf{I}^J \otimes  \textbf{Q}_K \otimes \textbf{I}^{P_J}
        \end{bmatrix}
,
\textbf{A} = \textbf{I}^{K}\otimes\begin{bmatrix}
        \textbf{A}_1 \\
        \vdots\\
        \textbf{A}_J \\
        \end{bmatrix}
\]

\noindent In the above, $j$ and $k$ denote the index of energy channel and basis material, respectively. $\textbf{A}_j$ and $\textbf{S}_j$ defines the system geometry and spectrum, respectively. $\textbf{B}_j$ encompasses the system blurring effects, including the focal-spot blur\cite{steven2016modeling}, gantry-rotation blur\cite{steven2018high}, and detector blur\cite{tilley2017penalized}. Covariance matrix $\textbf{K}_j$ characterizes the measurements noise. 

\begin{table}[h]
\caption{Spectral CT System Parameters and Notation}
\label{tab:system}
\centering
\begin{tabular}{c|c|c}
\hline
Parameter  &   Notation    &   Size \\
\hline
Number of spectral channel   &  $J$ &\\
Number of basis material   &  $K$ &\\
Number of image voxel &  $M$ &  \\
Number of measurements of the $j$th channel&  $P_j$&  \\
Number of energy bins & $R$ & \\
The $k$th basis material image &  $\textbf{x}_k$ & $M\times1$\\
The $k$th basis material mass attenuation &  $\textbf{Q}_k$ & $R\times 1$\\
The $j$th channel system matrix &  $\textbf{A}_j$ &  $P_j \times M$\\
The $j$th channel spectrum &  $\textbf{S}_j$ &  $R \times 1 $ \\
The $j$th channel blur &  $\textbf{B}_j$ &  $P_j \times P_j $ \\
The $j$th channel measurements &  $\textbf{y}_j$ &  $P_j \times 1 $ \\
The $j$th channel measurement covariance &  $\textbf{K}_j$ &  $P_j \times P_j $ \\
\hline
\end{tabular}
\end{table}

The forward model \eqref{eq:phyics_model} accommodates any number of basis materials and spectral channels. Additionally, it accounts for arbitrary channel-wise spectra, system geometries, and the blur models, permitting accurate representations of a wide range of spectral systems including photon-counting\cite{willemink2018photon}, dual-layer\cite{rassouli2017detector}, and kV-switching\cite{mahmood2018rapid} CT as well as more exotic spectral CT systems that utilize unusual/sparse sampling approaches\cite{tivnan2022design}. Material decomposition aims to reconstruct the basis material density $\textbf{x}$ from the spectral measurements $\textbf{y}$. Model-based algorithms perform the material decomposition by maximizing the posterior probability:

\begin{equation}
\begin{split}
    \hat{\textbf{x}} &= \text{argmax} \log p(\textbf{x}|\textbf{y}) \\
    & = \text{argmax} \log p(\textbf{y}|\textbf{x}) +  \log p(\textbf{x}) \\
    & = \text{argmax} \underbrace{-\|\textbf{BS}\exp({-\textbf{QAx}})-\textbf{y}\|_{\textbf{K}^{-1}}^2}_{Likelihood} + \underbrace{\log p(\textbf{x})}_{Prior}
    \label{eq:MAP}
\end{split}
\end{equation}
\noindent The physical model of the spectral system determines the likelihood term, while the prior information of $\textbf{x}$ is encoded in the second term. Mathematically, \eqref{eq:MAP} is a nonlinear problem without an explicit solution, therefore solutions are typically estimated through iterative optimization algorithms.

\begin{figure*}[h]
\centering
\includegraphics[width=\textwidth]{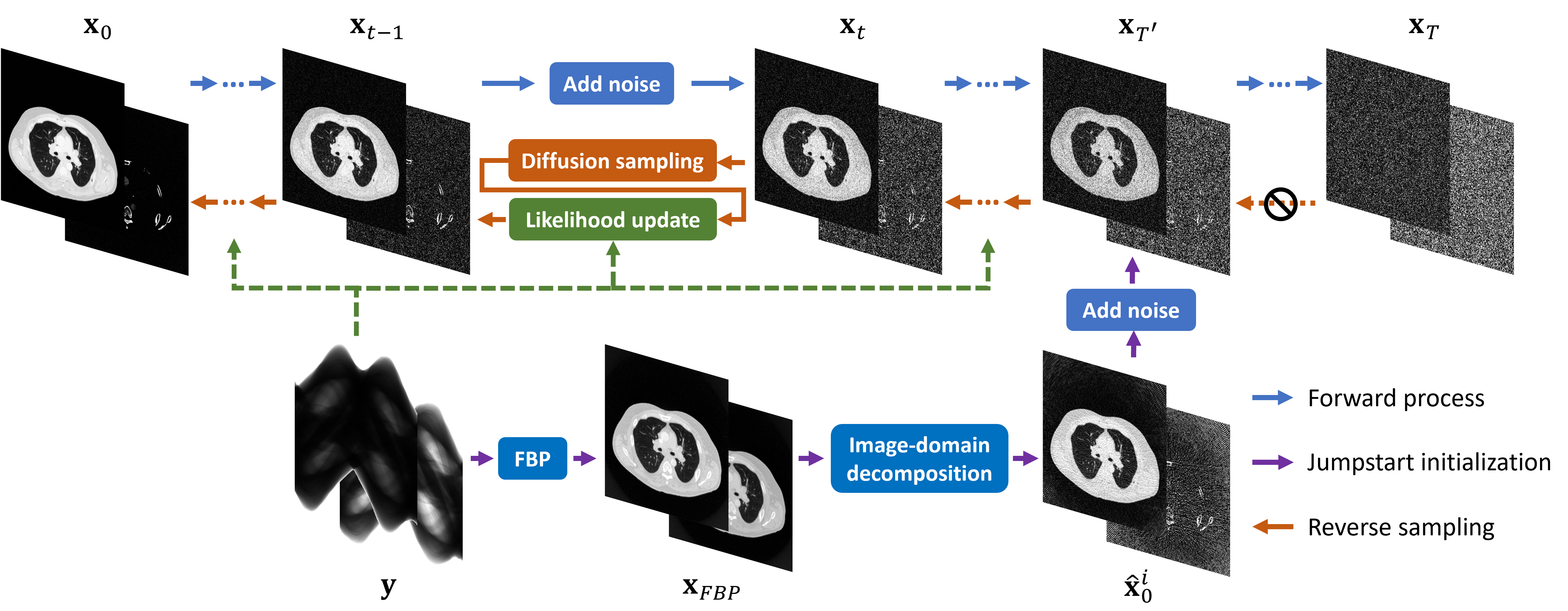}
\centering
\caption{Workflow of spectral diffusion posterior sampling. In the forward process a set of material basis images is simultaneously and incrementally degraded by noise. A reverse process is defined that restores those images conditioned on the measurement data. The jumpstart procedure is also illustrated permitting one to skip a number of initial reverse time steps for computation reduction and improved stability.}
\label{fig:dps} 
\end{figure*}

\subsection{Spectral Diffusion Posterior Sampling}
\label{sec:dps}
\subsubsection{Diffusion Model}
Conventional model-based algorithms commonly apply relatively simple analytic forms for the prior distribution, which typically do not capture the complexity of the actual prior distribution. Generative models \cite{ruthotto2021introduction} have been widely explored to accurately represent high-dimensional data distributions. Diffusion models\cite{ho2020denoising} have became one of the most popular generative models and provide a strategy to learn the data distribution by estimating the gradient of the prior probability density $\nabla_{\textbf{x}}\log p_t(\textbf{x})$. Specifically, given a target-domain dataset, a diffusion model gradually degrades each sample with time-dependent Gaussian noise described by a forward stochastic differential equation (SDE):
\begin{equation}
\label{eq:forward}
    \mathrm{d}\textbf{x} = -\frac{\beta_t}{2} \textbf{x}\mathrm{d}t + \sqrt{\beta_t}\mathrm{d}\textbf{w}, 
\end{equation}

\noindent where $\beta_t$ defines the time-dependent noise variance and $\mathrm{d}\textbf{w}$ represents the standard Wiener process. This forward process may be inverted through a reverse SDE:
\begin{equation}
\label{eq:reverse}
    \mathrm{d}\textbf{x} = [-\frac{\beta_t}{2} \textbf{x}-\beta_t\nabla_{\textbf{x}_t} \mathrm{log} p_t({\textbf{x}_t})]\mathrm{d}t + \sqrt{\beta_t}\mathrm{d}\textbf{w}.
\end{equation}

\noindent The forward process Eq.\eqref{eq:forward} gradually corrupts the sample from target distribution to pure noise. Conversely, the reverse process Eq.\eqref{eq:reverse} generates samples of the prior distribution starting from pure noise. The unknown probability gradient $\nabla_{\textbf{x}_t} \mathrm{log} p_t({\textbf{x}_t})$, referred to as the score function, may be approximated by a neural network trained on the target-domain dataset\cite{song2020score}.

\subsubsection{Diffusion Posterior Sampling}
Diffusion posterior sampling (DPS) applies the diffusion model to permit sampling from the posterior probability $p({\textbf{x}|\textbf{y}})$. The reverse process of DPS is written as:
\begin{equation}
\label{eq:dps}
\begin{split}
    \mathrm{d}\textbf{x} = &[-\frac{\beta_t}{2} \textbf{x}-\beta_t\nabla_{\textbf{x}_t} \mathrm{log} p_t({\textbf{x}_t|\textbf{y}})]\mathrm{d}t + \sqrt{\beta_t}\mathrm{d}\textbf{w}\\
    = &\underbrace{[-\frac{\beta_t}{2} \textbf{x}-\beta_t\nabla_{\textbf{x}_t} \mathrm{log} p_t({\textbf{x}_t})]\mathrm{d}t + \sqrt{\beta_t}\mathrm{d}\textbf{w}}_{Unconitional \ reverse \ sampling, \ Eq.(4)} \\
    &- \underbrace{\beta_t\nabla_{\textbf{x}_t} \mathrm{log} p_t({\textbf{y}|\textbf{x}_t})\mathrm{d}t}_{Likelihood \ gradient}.
\end{split}
\end{equation}

\noindent Eq.\eqref{eq:dps} illustrates that the posterior reverse sampling may be split into unconditional reverse sampling and gradient descent on the data likelihood. The intractable likelihood on $\textbf{x}_t$ is further approximated using Tweedie’s formula\cite{efron2011tweedie}:
\begin{equation}
\label{eq:dps_appro}
    p_t(\textbf{y}|\textbf{x}_t) \approx p_t(\textbf{y}|\hat{\textbf{x}}_0), \text{where} \ \hat{\textbf{x}}_0 = 
\frac{1}{\sqrt{\overline{\alpha}_t}}(\textbf{x}_t+(1-\overline{\alpha}_t))\textbf{s}_\theta(\textbf{x}_t,t).
\end{equation}

In the context of medical imaging reconstruction, the data likelihood is formed based on a physical model of the measurements $\textbf{y}$, and in Eq.\eqref{eq:dps}, the data likelihood is separate from the unconditional reverse sampling, which allows one to train a purely generative model on the target-domain dataset then apply it to posterior sampling on arbitrary imaging systems. This sampling strategy provides a flexible framework to incorporate an accurate physics model and sophisticated prior information into the image reconstruction.

\subsubsection{Material Decomposition Using Spectral DPS}
Combining the spectral CT physics model and diffusion posterior sampling, we have DPS for multi-material reconstruction and decomposition:
\begin{equation}
\label{eq:dps_spectral}
\begin{aligned}
    \mathrm{d}\textbf{x} = &[-\frac{\beta_t}{2} \textbf{x}-\beta_t\nabla_{\textbf{x}_t} \mathrm{log} p_t({\textbf{x}_t})]\mathrm{d}t + \sqrt{\beta_t}\mathrm{d}\textbf{w} \\
    &+ \beta_t \nabla_{\textbf{x}_t} \left\| \textbf{BS}\exp({-\textbf{A}\hat{\textbf{x}}_0}) -\textbf{y}\right\|_{\textbf{K}^{-1}}^2\mathrm{d}t
\end{aligned}
\end{equation}

\noindent For the training stage, material basis images are concatenated as a multi-channel input, and independent Gaussian noise is added to each channel individually then jointly predicted by a single neural network. We refer to the sampling \eqref{eq:dps_spectral} as \textbf{Baseline DPS}. While baseline DPS has demonstrated the ability to produce realistic images\cite{li2024ctreconstructionusingdiffusion}, such an approach can suffer low sampling efficiency, large sampling variability, and poor data consistency, particularly for ill-conditioned problems such as one-step material decomposition. To address these problems, this work employs several strategies developed in our previous work:
\begin{itemize}[leftmargin=*]
    \item A jumpstart strategy is applied wherein many reverse time steps are skipped. For many medical imaging reconstruction problems, a direct first-pass reconstruction $\hat{\textbf{x}}_0$ can be efficiently computed. Though the first-pass results may be corrupted by noise and various artifacts, this discrepancy with ground truth $\textbf{x}_0$ is mitigated through the forward diffusion since the random noise gradually dominates the images. For sufficiently large $t=T'$, $\textbf{x}_{T'}$ and $\hat{\textbf{x}}_{T'}$ conform to almost the same distribution. This permits starting reverse sampling from $\hat{\textbf{x}}_{T'}$ rather than the pure noise. In this work, filtered backprojection followed by image-domain decomposition (detailed in the Sec.\ref{sec:comp}), is used to approximate $\hat{\textbf{x}}_0$, and the optimal $T'$, which controls the amount of noise added to $\hat{\textbf{x}}_0$, is determined by a parameter sweep. 

    \item The computation of the likelihood gradient with respect to $\textbf{x}_t$ is both memory and time intensive due to the propagation through the deep network \cite{jiang2024strategies}. Here we adopt an approach that computes the gradient with respect to $\hat{\textbf{x}}_0$ to reduce the computational burden as well as stabilize the gradient computation. In this case, the reverse sampling is written as:
    \begin{equation}
    \label{eq:spectraldps}
    \begin{aligned}
        \mathrm{d}\textbf{x} = &[-\frac{\beta_t}{2} \textbf{x}-\beta_t\nabla_{\textbf{x}_t} \mathrm{log} p_t({\textbf{x}_t})]\mathrm{d}t + \sqrt{\beta_t}\mathrm{d}\textbf{w} \\
        &+ \beta_t \nabla_{\hat{\textbf{x}}_0} \left\| \textbf{BS}\exp({-\textbf{A}\hat{\textbf{x}}_0}) -\textbf{y}\right\|_{\textbf{K}^{-1}}^2\mathrm{d}t
    \end{aligned}
    \end{equation}    
    
    \item The ordered subsets strategy\cite{erdogan1999ordered} is commonly applied in tomographic reconstruction for computational advantage. This technique is built on the principle that the likelihood gradient can be well approximated by the gradient computed on a subset of the whole projection set:
    \begin{equation}
    \label{eq:os}
        \nabla_{\textbf{x}_0}\mathcal{L} (\textbf{x}_0,\textbf{y}) \approx S\nabla_{\textbf{x}_0}\mathcal{L} (\textbf{x}_0,\textbf{y}_s),
    \end{equation}
    where $S$ and $s$ is the number of subsets and the subset index, respectively. In this work, The likelihood gradient is computed by using a custom-written auto-differentiable polychromatic forward projector, and the likelihood update applies all subset updates in each time step. As such, the number of subsets is the same as the number of gradient descent operations per time step. As in the previous work\cite{jiang2024strategies}, Adaptive Moment Estimation (Adam) optimizer is implemented to further improve the convergence rate with a smooth optimization trajectory and adaptive step size. 
\end{itemize}

We refer to the DPS equipped with these improved strategies as \textbf{Spectral DPS}. Implementation of spectral DPS is summarized in Fig.\ref{fig:dps} and with pseudocode in Algorithm 1. One of the distinct features of DPS, as compared to conventional algorithms, is that the results exhibit variability, i.e., sampling results vary even when conditioned on the same measurement realization due to the inherently stochastic processing. To optimize performance, we sweep three hyperparameters - the number of jumpstart time steps $T'$, the number of subsets $S$, and the step size $\eta$ to minimize variability of the decomposition results.

\begin{algorithm}
\caption{Spectral DPS}
\begin{algorithmic}[1]\small
\State $T$: diffusion training steps
\State $T'$: jumpstart steps, $T'\ll T$
\State $\hat{\textbf{x}}_0^i$: image-domain decomposition
\State
\State \# Adam optimizer 
\State $\eta$: step size
\State $\gamma_1 = 0.9, \gamma_2 = 0.999$: momentum coefficients 
\State
\State \# Order subset
\State $S$: number of subset
\State $\{\textbf{A}_s, \textbf{y}_s\}$: subset forward projector and spectral measurements
\State
\State \# Reverse sampling initialization
\State $\boldsymbol{\epsilon}\sim\mathcal{N}~(0,\textbf{I}^M\otimes \textbf{I}^K)$
\State $\textbf{x}_{T'} = \hat{\textbf{x}}_0^{T'}=\sqrt{\bar{\alpha}_{T'}}\hat{\textbf{x}}_0^i + \sqrt{1-\bar{\alpha}_{T'}}\boldsymbol{\epsilon}$
\State
\State \# Diffusion Posterior Sampling
\For{\texttt{$t = T'$ to $1$}}:
    \State \# Diffusion sampling:
    \State $\textbf{z} \sim \mathcal{N}~(0,\textbf{I}^M\otimes \textbf{I}^K)$
    \State $\hat{\textbf{x}}_0 = \frac{1}{\sqrt{\bar{\alpha}_t}}(\textbf{x}_t-\sqrt{1-\bar{\alpha}_t}\boldsymbol{\epsilon}_\theta(\textbf{x}_t,t))$
    \State $\textbf{x}_{t-1}' = \frac{\sqrt{\alpha_t}(1 - \bar{\alpha}_{t-1})}{1 - \bar{\alpha}_t} \textbf{x}_t + 
    \frac{\sqrt{\bar{\alpha}_{t-1}}\beta_t} {1 - \bar{\alpha}_t} \hat{\textbf{x}}_0 + \sigma_t \textbf{z}$
    \State
    \State \# Likelihood update:
    \State $\hat{\textbf{x}}_0' = \hat{\textbf{x}}_0$ 
    \For{\texttt{$s = 1$ to $S$}}:
        \State $\hat{\textbf{x}}_0' = \text{Adam}(\hat{\textbf{x}}_0',S\nabla_{\hat{\textbf{x}}_0'}\left\| \textbf{BS}\exp({-\textbf{A}_s\hat{\textbf{x}}_0'}) -\textbf{y}_s\right\|_{\textbf{K}^{-1}}^2)$
    \EndFor
    \State $\textbf{x}_{t-1} = \textbf{x}_{t-1}' - \hat{\textbf{x}}_0 + \hat{\textbf{x}}_0'$
\EndFor
\end{algorithmic}
\end{algorithm}

\subsection{Unconditional Diffusion Model Training}
\label{sec:train}
\subsubsection{Dataset Generation}
\begin{figure}[h]
\centering
\includegraphics[width=\linewidth]{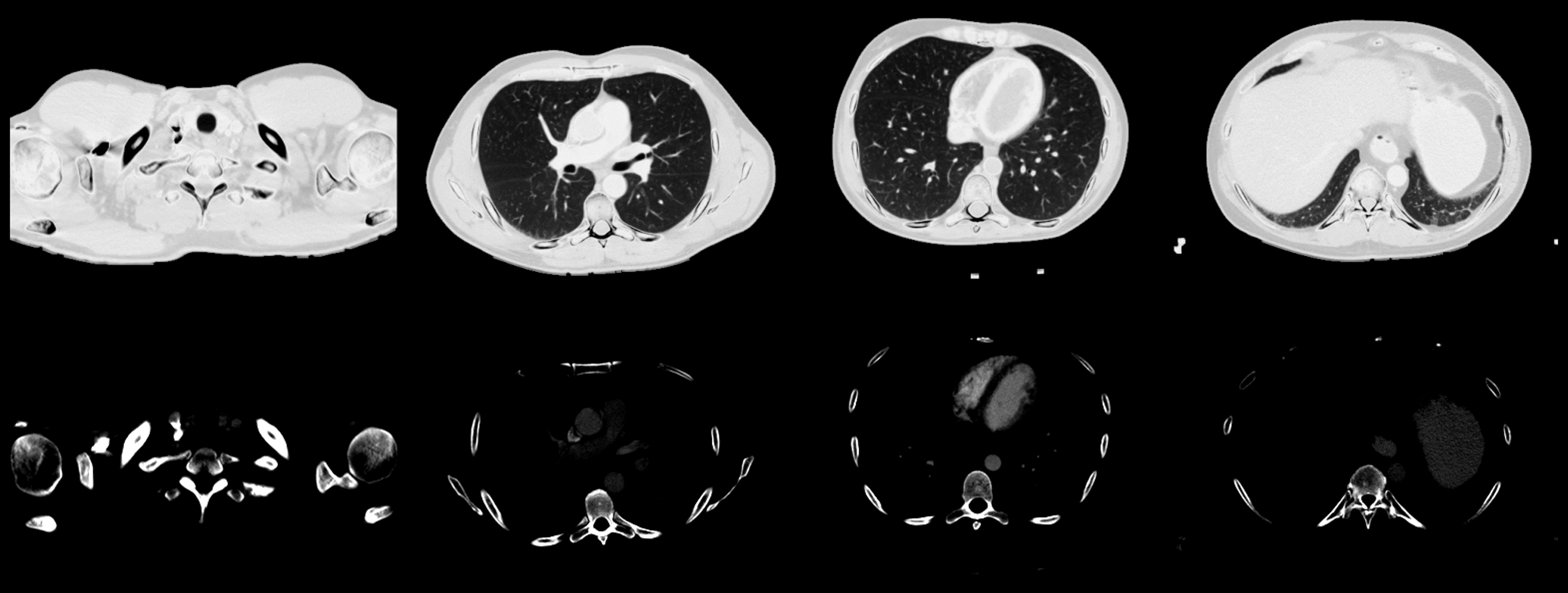}
\caption{Example images in the synthetic two-material dataset. Top and bottom rows are paired water and calcium images. Display window: water: $[0,1.0]g/ml$, calcium: $[0,0.2]g/ml$.}
\label{fig:example} 
\end{figure}

The dataset generation procedure is summarized in Fig.\ref{fig:workflow}. We investigated water/calcium decomposition in this work. A two-material dataset was created based on the public CT Lymph Nodes dataset\cite{roth2014new}, from which chest-region slices were extracted from 150 distinct patients, forming a CT dataset of 30000 slices. Pre-processing includes bed removal via morphological operations and metal reduction via clipping the maximum value to 2000HU. Hounsfield unit values were converted into attenuation coefficients, and then water and calcium densities (unit: $g/cm^3$) were determined by two soft-threshold functions:

\begin{figure}[t]
\centering
\includegraphics[width=\linewidth]{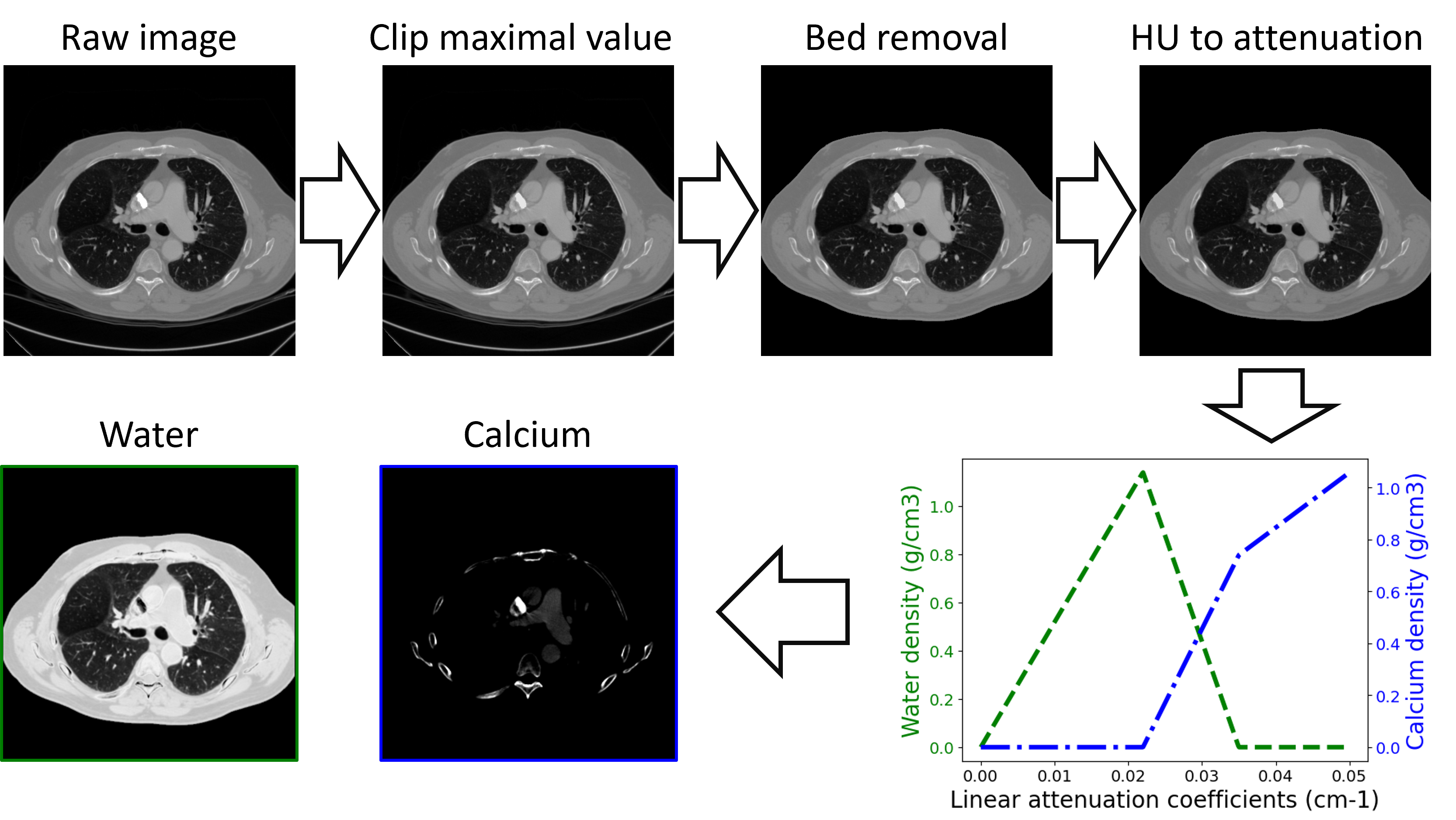}
\caption{Summary of the processing used to create the two-material dataset from public single-energy CT images.}
\label{fig:workflow} 
\end{figure}

\begin{subequations}
\label{eq:material}
\begin{equation}  
 x_{w}(\mu) = 
\begin{cases}
    k_w\mu, & \text{if } \mu \leq \mu_w \\
    k_w\mu_w - k_{wc}(\mu-\mu_w), & \text{if } \mu_w < \mu < \mu_c\\
    0, & \text{otherwise}
\end{cases} 
\end{equation}
\begin{equation}  
 x_{c}(\mu) = 
\begin{cases}
    k_c(\mu-\mu_c)+k_{cw}(\mu_c-\mu_w), & \text{if } \mu \geq \mu_c \\
    k_{cw}(\mu-\mu_w), & \text{if } \mu_w < \mu < \mu_c\\
    0, & \text{otherwise}
\end{cases}    
\end{equation}
\end{subequations}
\noindent The parameters, $k_w$, $k_{wc}$, $k_{cw}$, $k_{c}$ were empirically set to $5.18$, $-8.77$, $5.69$, $2.12~g/cm^2$, while $\mu_w$ and $\mu_c$ were $0.22~cm^{-1}$ and $0.35~cm^{-1}$. Fig.\ref{fig:example} illustrates example images in the synthetic two-material dataset. Part of the soft tissue, notably in the heart region, exhibits high intensity in the calcium images because of iodine contrast enhancement.

\subsubsection{Network Training}
A Residual Unet\cite{zhang2018road} was employed as the backbone network of diffusion model. Seven encoder/decoder blocks were configured and the number of channels was set to ($16,32,64,128,256,512,512$). Paired water and calcium images were concatenated to form the 2-channel input $\textbf{x}_0$. The diffusion process was discretized into $T = 1000$ time steps, with the variance of added noise linearly increasing from $\beta_1 = 1e^{-4}$ to $\beta_{1000} = 0.02$. 25000 slices of the processed images are used for model
training, and the slices from patients excluded in the training dataset are reserved for evaluation. The model was implemented in the PyTorch framework and was optimized by Adam optimizer with a batch size of $32$ and a learning rate of $10^{-4}$. Training stopped after $200$ epochs.

\subsection{Evaluation}
\label{sec:eval}
\begin{figure}[h]
\centering
\includegraphics[width=\linewidth]{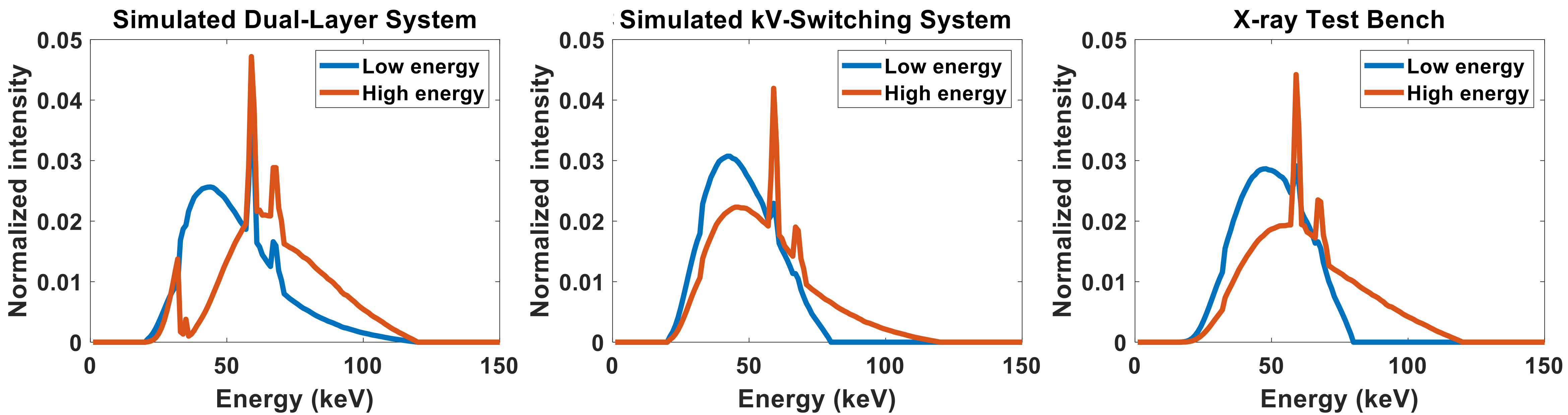}
\caption{Normalized spectral sensitivities of the CT systems used in the simulation studies as well as the estimated sensitivities for the physical data experiments.}
\label{fig:spectrum} 
\end{figure}

\subsubsection{Simulation Study}
We investigated the proposed material decomposition for both dual-layer CT and kV-switching CT systems in a simulation study. Each system employs $720$ projections with Poisson noise equivalent to an exposure of $0.1$mAs/view. The source-to-detector distance (SDD) was $1000$mm, and source-to-axis distance (SAD) $500$mm. System blur was not modeled. Reconstruction voxel size and detector pixel size was set to $0.8$mm and $1.0$mm, respectively. The simulated kV-switching system alternates the tube voltage between $80$kV and $120$kV every other view. In the dual-layer system, the mismatch geometry of the two channels was modeled with a $5$mm gap between the two layers. Fig.\ref{fig:spectrum} displays the normalized spectrum sensitivities of the investigated systems.

\subsubsection{Physical Phantom Study}
\begin{figure}[h]
\centering
\includegraphics[width=\linewidth]{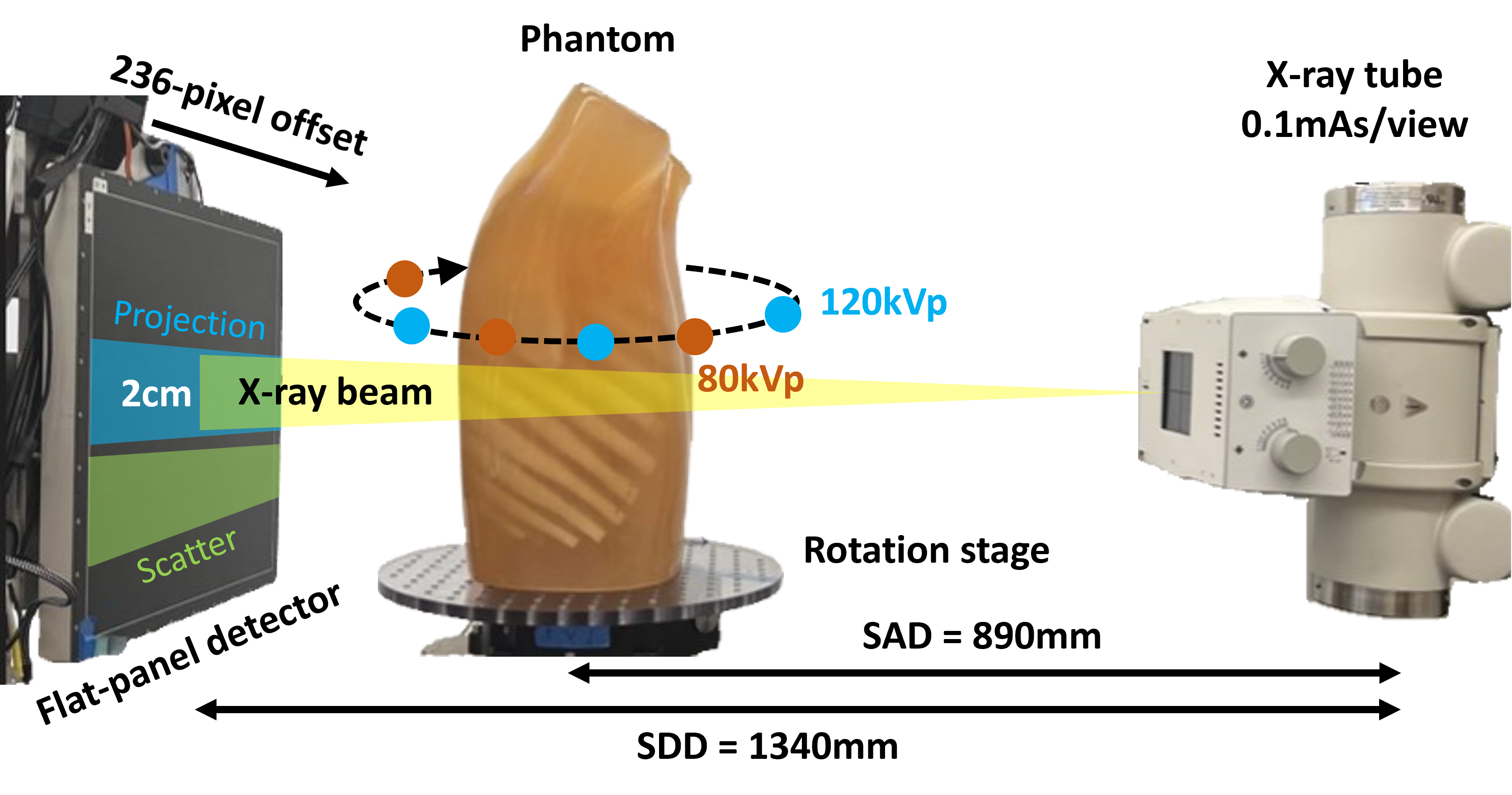}
\caption{Illustration of the benchtop cone-beam CT system used in physical experiments.}
\label{fig:hardware} 
\end{figure}

A phantom study was conducted by scanning an anthropomorphic lung phantom (PH-1, Kyoto Kagaku, Japan) on a cone-beam CT (CBCT) test bench system (Fig.\ref{fig:hardware}). Two sets of projections ($720$views/set) were acquired at 80kV and 120kV with an exposure of $0.1$mAs/view, which were then merged to mimic a dual-energy scan. The x-ray beam was vertically collimated to $2$cm width on the detector to minimize scatter, and the signal behind the collimator was used to estimate residual scatter, which was further subtracted from the raw projections. 

\subsubsection{Evaluation Metrics}
Image STD was computed to evaluate the sampling variability:
\begin{equation}
\text{STD} = \|\sqrt{\mathbb{E}\{(\mathbb{E}\{\hat{\textbf{x}}\}-\hat{\textbf{x}})^2}\}\|_1,
\end{equation}
\noindent and image bias was computed to quantify the numerical accuracy:
\begin{equation}
\text{Bias} = \|\mathbb{E}\{\hat{\textbf{x}}\}-\textbf{x}\|_1.
\end{equation}
\noindent where the expectation was evaluated on 32 DPS samples from a common set of measurements. RMSE, PSNR, and SSIM was also computed to evaluate the image quality and explore the relationship between variability and other metrics:
\begin{subequations}
\begin{equation}
\text{RMSE} = \mathbb{E}\{\sqrt{MSE(\hat{\textbf{x}},\textbf{x})}\}
\end{equation}
\begin{equation}
\text{PSNR} = \mathbb{E}\left\{10 \cdot \log_{10} \left( \frac{MAX_\textbf{x}^2}{MSE(\hat{\textbf{x}},\textbf{x})} \right) \right\} 
\end{equation}
\begin{equation}
\text{SSIM} = \mathbb{E}\left\{\frac{(2\mu_{\hat{\textbf{x}}}\mu_\textbf{x} + c_1)(2\sigma_{\hat{\textbf{x}}\textbf{x}} + c_2)}{(\mu_{\hat{\textbf{x}}}^2 + \mu_\textbf{x}^2 + c_1)(\sigma_{\hat{\textbf{x}}}^2 + \sigma_\textbf{x}^2 + c_2)}\right\}.
\end{equation}
\end{subequations}

\noindent In a comparison study, we investigated decomposition results from different algorithms and focus on region-of-interests (ROIs) containing fine structures. The RMSE, PSNR, and SSIM are computed within each ROI to demonstrate decomposition performance. 

\subsubsection{Comparison Methods and Parameter Selection}
\label{sec:comp}
Image-Domain Material Decomposition (IDD): Filtered backprojection (FBP) was first applied to reconstruct each energy channel, then a pixel-wise matrix inversion was utilized to decompose the attenuation maps into density maps, this inversion may be represented as:  

\begin{equation}
\label{eq:IDD}
\begin{bmatrix}
x_1 \\
\vdots\\
x_K \\
\end{bmatrix} =
\textbf{M}^{-1}
\begin{bmatrix}
\mu_1 \\
\vdots\\
\mu_J \\
\end{bmatrix}
\end{equation}

\noindent where $[\textbf{M}]_{jk}=\mu_{jk}$ is the effective mass attenuation of the $k$th basis material for the $j$th spectral channel. In this work, the number of bases and spectral channels is equal with $K=J=2$, and the $2\times2$ matrix $\textbf{M}^{-1}$ is pre-calibrated using 200 randomly selected slices in the training dataset through least-square minimization. 

Model-based Material Decomposition (MBMD): We follow Tilley et.al.\cite{tilley2019model} to implement one-step model-based decomposition. The objective function is composed of data likelihood and quadratic penalty terms:
\begin{equation}  
\label{eq:mbmd}
\underbrace{\|\textbf{BS}\exp({-\textbf{QAx}})-\textbf{y}\|_{\textbf{K}^{-1}}^2}_{Likelihood}+\underbrace{\lambda_{water}\textbf{R}(\textbf{x}_{water})+\lambda_{Ca}\textbf{R}(\textbf{x}_{Ca})}_{Penalty}
\end{equation}

\noindent This objective is minimized using a Separable Paraboloidal Surrogate (SPS) technique to form the water and calcium density maps. In this work, we used 8000 iterations to obtain a reasonably converged solution. The material-specific regularization strength was tuned using a 2D sweep for $\lambda_{water}$ and $\lambda_{Ca}$. The optimal combination was selected as the one with minimal RMSE.

Baseline DPS: Baseline DPS was implemented as described in Ref.\cite{chung2022diffusion}. In essence, the jumpstart, ordered subsets, and Adam optimizer strategies from Algorithm 1 were removed. The step size of gradient descent was determined as suggested in \cite{li2024ctreconstructionusingdiffusion}:
\begin{equation}
    \eta_t = \eta/\|\nabla_{\textbf{x}_t}\log p(\textbf{y}|\textbf{x}_t)\|_2
\end{equation}
\noindent We swept the constant $\eta$ from $1$ to $100$, and selected the one with minimal STD.

\begin{figure}[h]
\centering
\includegraphics[width=\linewidth]{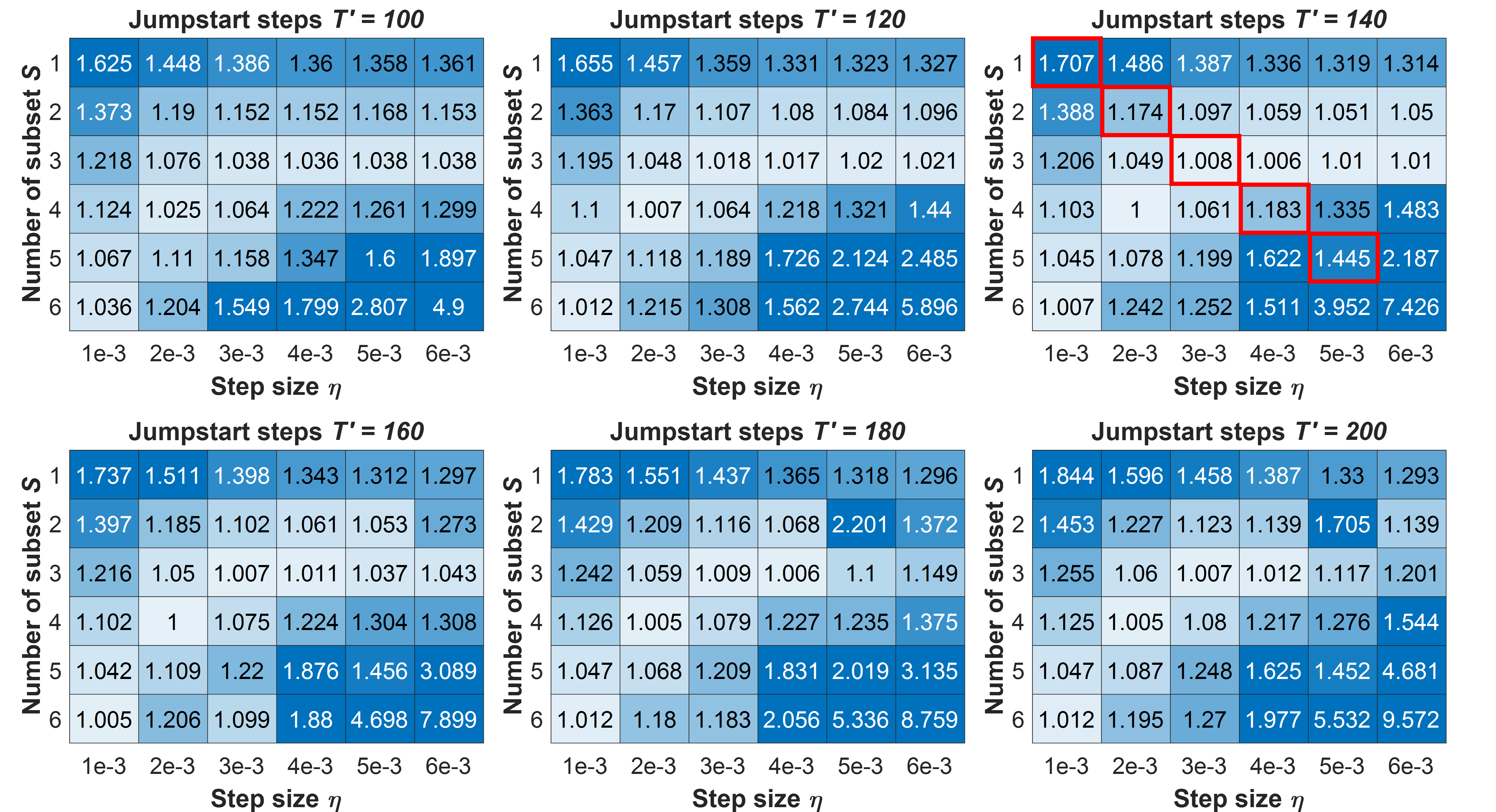}
\centering
\caption{STD maps for different combinations of hyperparameters. All values were normalized by the minimal STD so that the lowest variability is unity. Parameter settings explored in Fig.\ref{fig:stdana} are outlined with red boxes.}
\label{fig:stdmap} 
\end{figure}

\section{Results}
\label{sec:results}

\begin{figure*}[h]
\centering
\includegraphics[width=\textwidth]{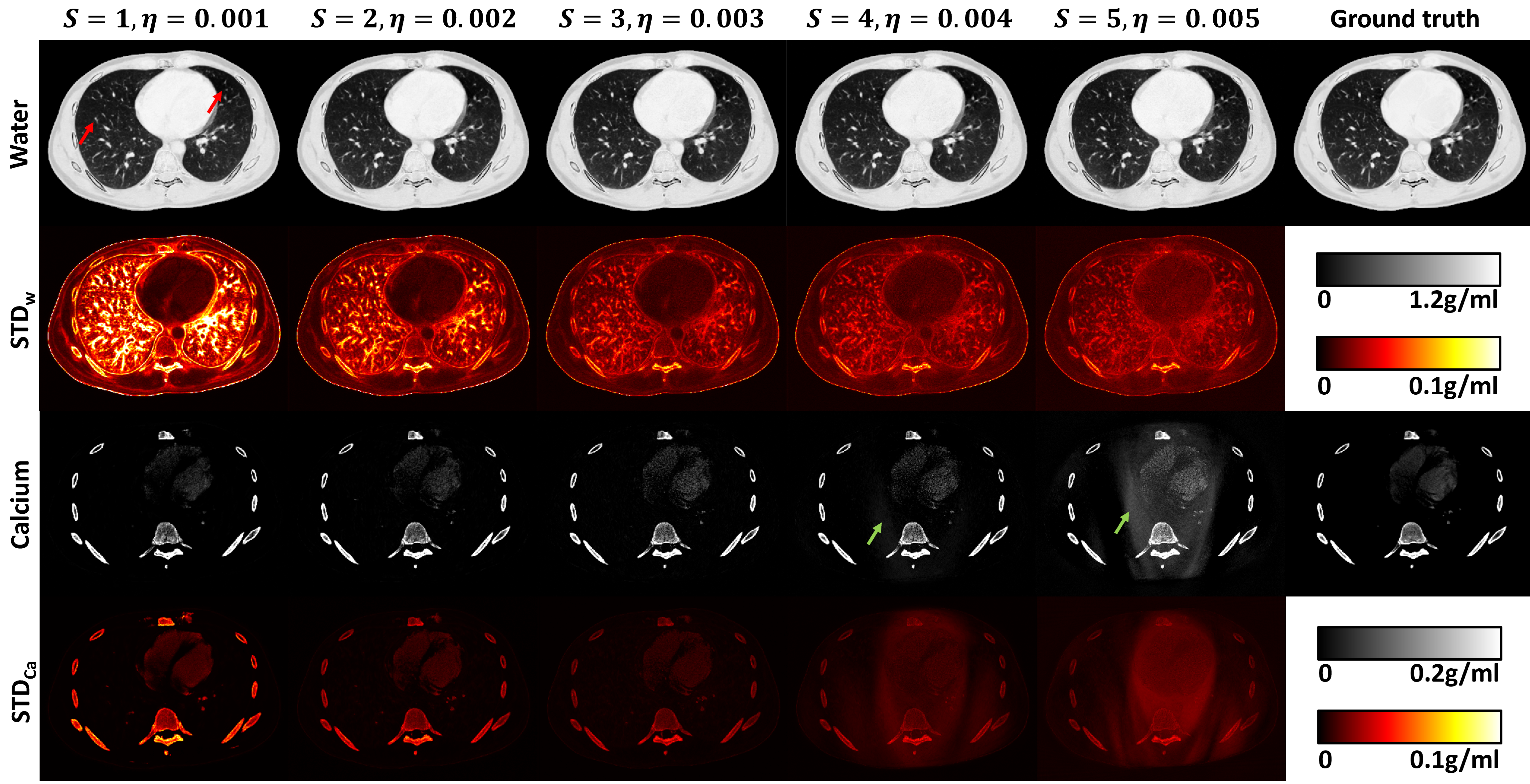}
\centering
\caption{Sample Spectral DPS results and the corresponding STD maps with different parameter settings. STD is computed on 32 DPS samples from the same set of measurement. Red arrows show false features created for low $S, \eta$ settings. The green arrow shows increased error for high $S, \eta$ in the background.}
\label{fig:stdana} 
\end{figure*}

\subsection{Parameter Optimization}
\begin{figure}[h]
\centering
\includegraphics[width=\linewidth]{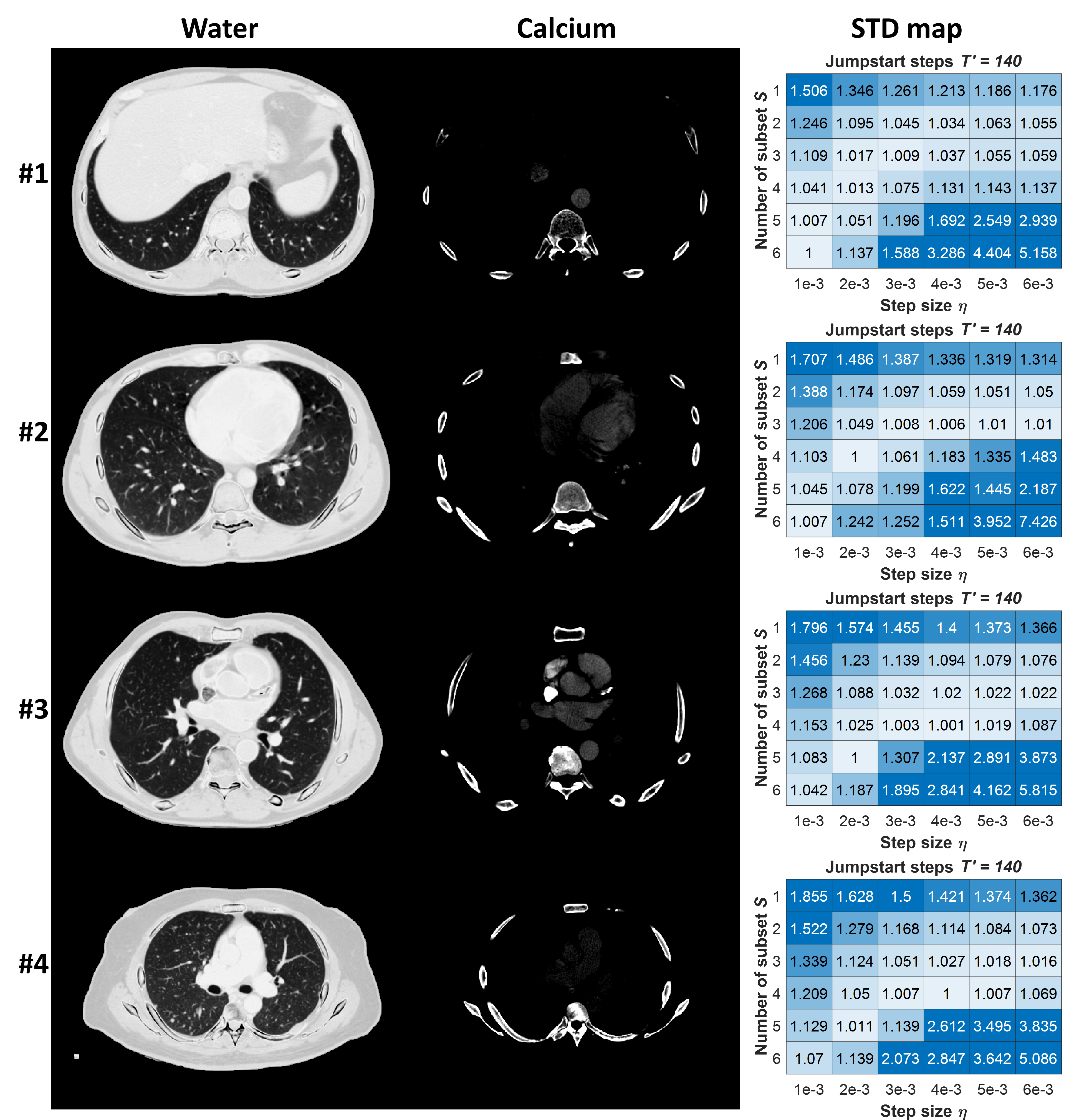}
\centering
\caption{Normalized STD map (right) for four slices (left) at different locations. Display window: water: $[0,1.2]g/ml$, calcium: $[0,0.2]g/ml$}
\label{fig:std_multi} 
\end{figure}
In this section, we present the results of the parameter sweep for the dual-layer CT system. 

\subsubsection{Variability Analysis}
The parameter sweep was performed for $T'$, $\eta$, and $S$, and the corresponding normalized STD heatmaps are displayed in Fig.\ref{fig:stdmap}. The subplots for each time step exhibit the similar trends. There is a low-variability region extending from top right to bottom left. For $T'\geq120$, the difference in the minimal STD for optimal $\eta, S$ is smaller than $<1\%$.

To investigate how the parameter $\eta, S$ selection affects the decomposition results, we display the decomposition and corresponding STD map of 5 different parameter settings with $T'=140$ in Fig.\ref{fig:stdana}. For small $S,\eta$ (e.g. $S=1,\eta=0.001$), the STD map shows large variability particularly around the edges, as well as features that, while realistic, are inconsistent with the ground truth. Increasing $S,\eta$ within a certain range led to significant STD reduction. However, when the optimal point was exceeded, i.e., $S=3,\eta=0.003$ in Fig.\ref{fig:stdana}, the STD increases again. Unlike the high STD caused by variation in fine structures, this increase was due to background signal fluctuations.

In Fig.\ref{fig:std_multi}, we plot the STD map of slices at different anatomical locations, revealing that the heatmaps of different slices displayed similar trends. This suggests that parameters optimized on one slice or a group of representative slices can be reasonably applied to other slices, thereby eliminating the need for an anatomical-specific parameter tuning (at least for these thoracic CT). This consistency across slices underscores the robustness and generalizability of the parameter optimization process, streamlining the application of spectral DPS.

\subsubsection{Relationship between STD and Other Metrics}
\begin{figure}[h]
\centering
\includegraphics[width=\linewidth]{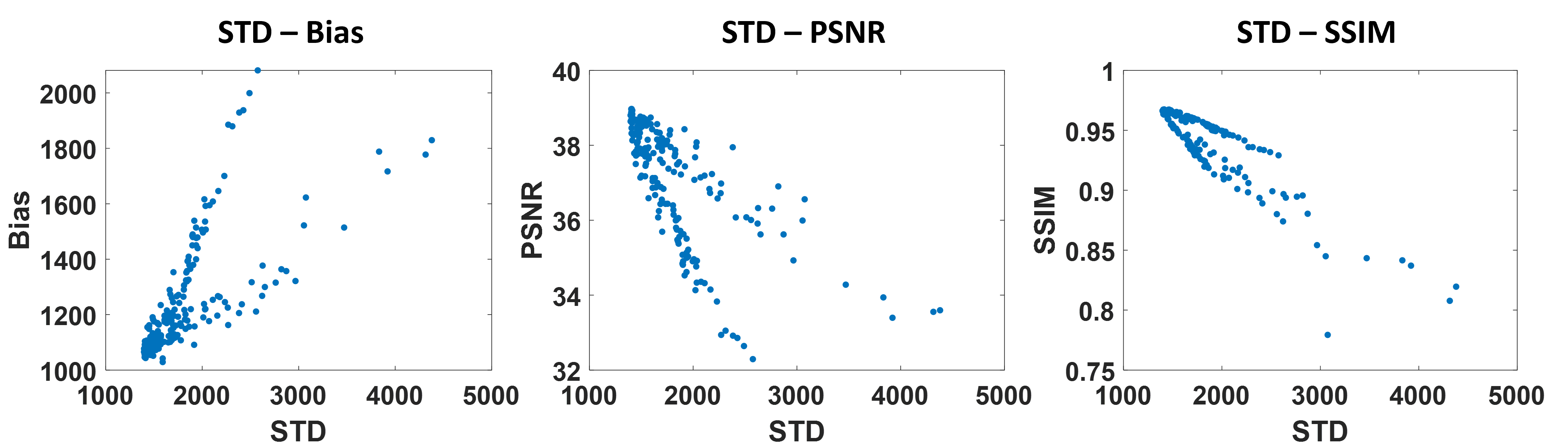}
\centering
\caption{Relationship between STD and bias/PSNR/SSIM. Note that minimum STD is highly correlated with optimal image quality metrics.}
\label{fig:std_metric} 
\end{figure}

Parameters optimized by minimizing STD alone reduces variability without taking other image properties into account. Fig.\ref{fig:std_metric} illustrates the relationship between STD and other metrics, i.e., bias, PSNR, and SSIM. Scatter plots between STD and the image quality metrics are formed by plotting these values over all cases in the 3D $T', \eta, S$ sweep. The scatter plots demonstrate that the settings with minimal STD also tend to produce images with minimal bias, maximal PSNR, and maximal SSIM. This indicates parameters optimization based on minimal STD also enhances decomposition accuracy and structural similarity. Based on this analysis, in this work, the hyperparameters of spectral DPS are determined by minimizing the decomposition STD, ensuring that the optimization process not only reduces variability but also improves overall image quality and consistency. While these trends are based on the dual-layer system, we find similar behavior for other dual-energy CT simulation.

\subsection{Simulation Study Results}
\begin{figure*}[h]
\centering
\includegraphics[width=\textwidth]{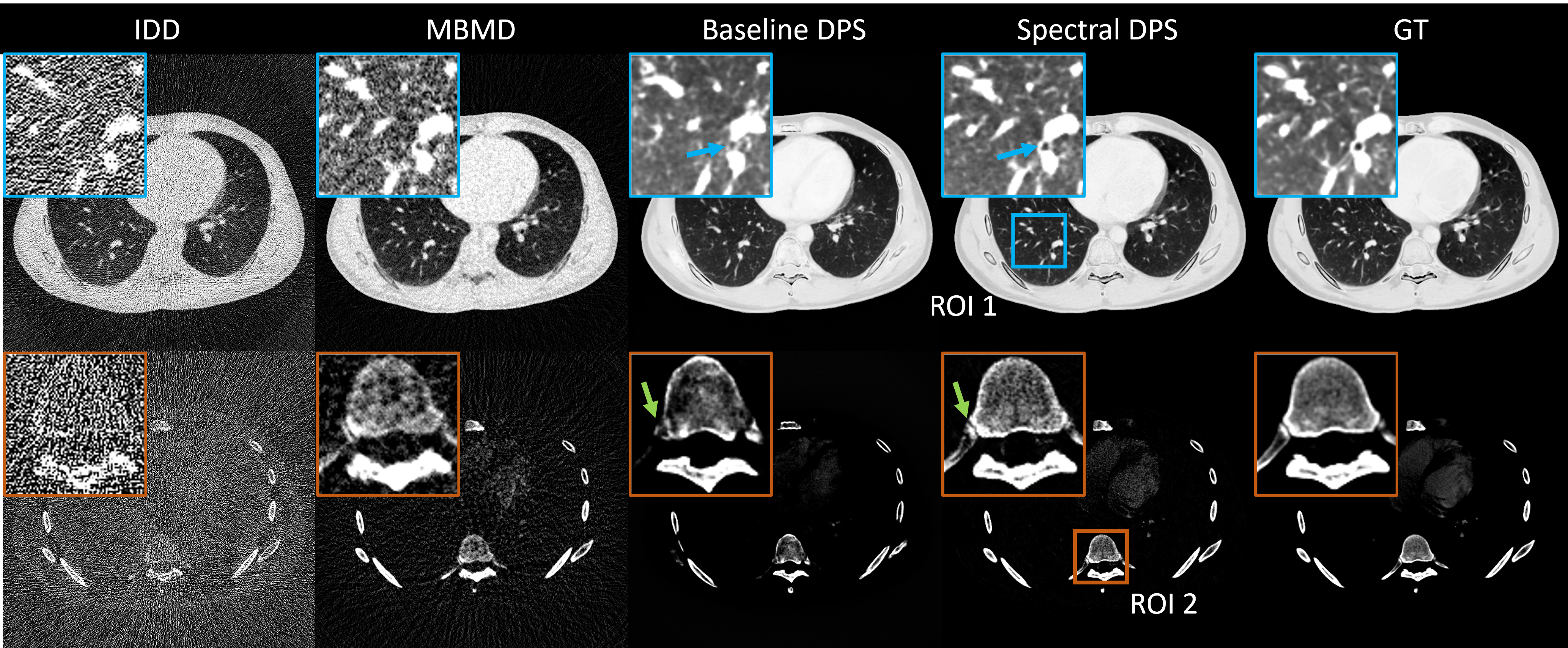}
\centering
\caption{Material decompositions for simulated dual-layer CT system. Display window: water(top): $[0,1.2]g/ml$, calcium(bottom): $[0,0.2]g/ml$.}
\label{fig:dual_layer} 
\end{figure*}

\begin{figure*}[h]
\centering
\includegraphics[width=\textwidth]{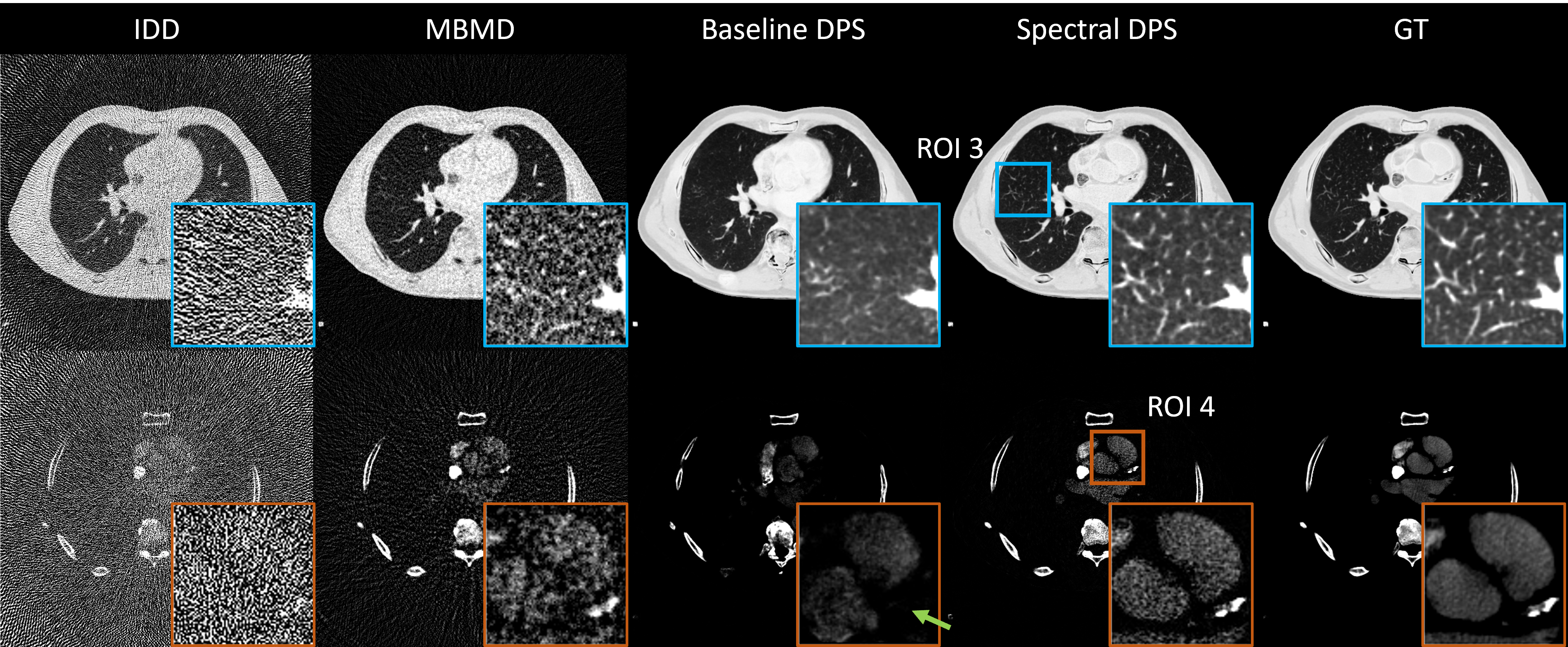}
\centering
\caption{Material decompositions for simulated kV-switching CT system. Display window: water(top): $[0,1.2]g/ml$, calcium(bottom): $[0,0.2]g/ml$.}
\label{fig:dual_kV} 
\end{figure*}

We apply spectral DPS to two different spectral CT systems and compare relative performance to alternate processing approaches.

\subsubsection{Qualitative Analysis}
A summary of material decompositions for dual-layer CT and kV-switching CT is shown in Fig.\ref{fig:dual_layer} and \ref{fig:dual_kV}, respectively. For the dual-layer system, IDD exhibited significant bias in both water and calcium images due to linear approximation of spectral effects. MBMD removed much of the background noise and improved the visibility of large features in the lung as well as the spine. Baseline DPS further suppressed the decomposition noise, generating realistic features in both material bases. However, significant deviations from ground truth were observed in the lung and spine. In particular, note a small airway (blue arrow) in lung ROI 1 and the spine boundaries (green arrow) in ROI 2 were misrepresented. Spectral DPS better recovers features and textures as compared with the ground truth, accurately depicting the lung airway (blue arrow) and the costotransverse joint (green arrow). 

The decomposition performance on the kV-switching system mirrored that of the dual-layer system. To illustrate performance in more challenging ROIs, we specified ROI 3 which contains many fine details within the lung parenchyma, and ROI 4 which contains low-density iodine-enhanced regions. The ROI 3 fine details were obscured by noise in IDD and difficult to visualize in MBMD, whereas spectral DPS faithfully recovered most of the features where the baseline DPS fails. The noise level in IDD and MBMD results are too high to reliably visualize the iodine-enhanced boundaries. Baseline DPS formed enhanced regions with somewhat accurate boundaries (green arrow) but fails to visualize the calcification on the bottom right. In contrast, spectral DPS not only reconstructed the overall shape of the enhanced region - albeit with slightly more noise than the ground truth - but also depicted the calcification with a distinct and clear boundary.

\subsubsection{Quantitative Evaluation}
Quantitative metrics computed for each ROI are summarized in TABLE.\ref{tab:metric}. Consistent across all ROIs, these metrics demonstrated similar trends. Baseline DPS achieves higher SSIM compared to MBMD, suggesting that it is more perceptually similar to the truth. In contrast, MBMD excelled in PSNR, indicating a superior capacity for imaging accuracy compared to baseline DPS. Spectral DPS demonstrated remarkable improvements on both metrics, enhancing SSIM  $63.82\%,55.01\%,87.76\%,31.47\%$ compared with baseline DPS, and  boosting PSNR by $39.87\%,27.94\%,57.30\%,26.53\%$ relative to MBMD. These significant improvements in quantitative metrics not only support the qualitative assessments but also underscore the superior imaging accuracy and robust performance that Spectral DPS offers across different spectral imaging systems.

\begin{table}[h]
\caption{Performance of different reconstruction algorithms on the simulated spectral CT system}
\label{tab:metric}
\centering
\begin{tabular}{|c|c|c|c|c|c|}
\hline
\multicolumn{1}{|c}{} & \multicolumn{1}{c|}{} & Algorithm & RMSE & SSIM & PSNR \\
\hhline{|=|=|=|=|=|=|}
\multirow{8}{*}{\rotatebox{90}{Dual-layer CT}} & \multirow{4}{*}{ROI 1}  & IDD & 0.3977 & 0.0113 & 9.1535 \\
                                               &  & MBMD & 0.0897 & 0.4778 & 22.0819 \\
                                               &  & Baseline DPS & 0.1066 & 0.5329 & 20.5847\\
                                               &  & Spectral DPS & \textbf{0.0325} & \textbf{0.8730} & \textbf{30.8850}\\
                                               \cline{2-6}
                                               & \multirow{4}{*}{ROI 2}  & IDD & 0.2219 & 0.0783 & 11.3898\\
                                               &  & MBMD & 0.0589 & 0.4406 & 22.9147 \\
                                               &  & Baseline DPS & 0.1156 & 0.5609 & 17.0516\\
                                               &  & Spectral DPS & \textbf{0.0281} & \textbf{0.8695} & \textbf{29.3179}\\
\hhline{|=|=|=|=|=|=|}
\multirow{8}{*}{\rotatebox{90}{kV-switching CT}} & \multirow{4}{*}{ROI 3}  & IDD & 0.4054 & 0.0576 & 8.9792\\
                                               &  & MBMD & 0.1101 & 0.3181 & 20.3027\\
                                               &  & Baseline DPS & 0.0833 & 0.4543 & 22.7208\\
                                               &  & Spectral DPS & \textbf{0.0288} & \textbf{0.8530} & \textbf{31.9368}\\
                                               \cline{2-6}
                                               & \multirow{4}{*}{ROI 4}  & IDD & 0.2166 & 0.0176 & 10.7589\\
                                               &  & MBMD & 0.0369 & 0.3024 & 26.1144\\
                                               &  & Baseline DPS & 0.0378 & 0.5700 & 25.9165\\
                                               &  & Spectral DPS & \textbf{0.0166} & \textbf{0.7494} & \textbf{33.0435}\\
\hline
\end{tabular}
\end{table}

\subsection{Physical Phantom Study Results}
\begin{figure*}[h]
\centering
\includegraphics[width=\textwidth]{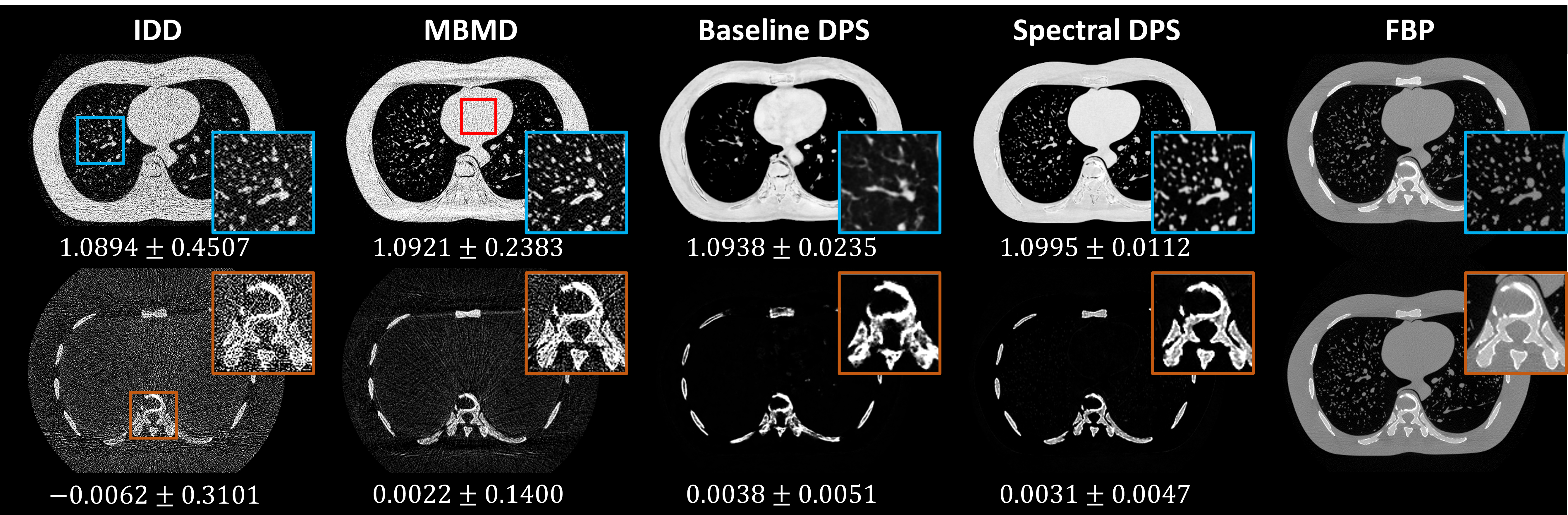}
\centering
\caption{Material decomposition on the bench CBCT system. The mean and STD computed on a homogeneous region (red box) and are listed below each material image. A single-energy FBP reconstruction is shown in the last column to provide a low-noise reference of the real phantom structure. Display window: water(top): $[0.2,1.2]g/ml$, calcium(bottom): $[0,0.4]g/ml$, FBP: $[0,0.04]mm^{-1}$.}
\label{fig:real} 
\end{figure*}

\begin{figure}[h]
\centering
\includegraphics[width=\linewidth]{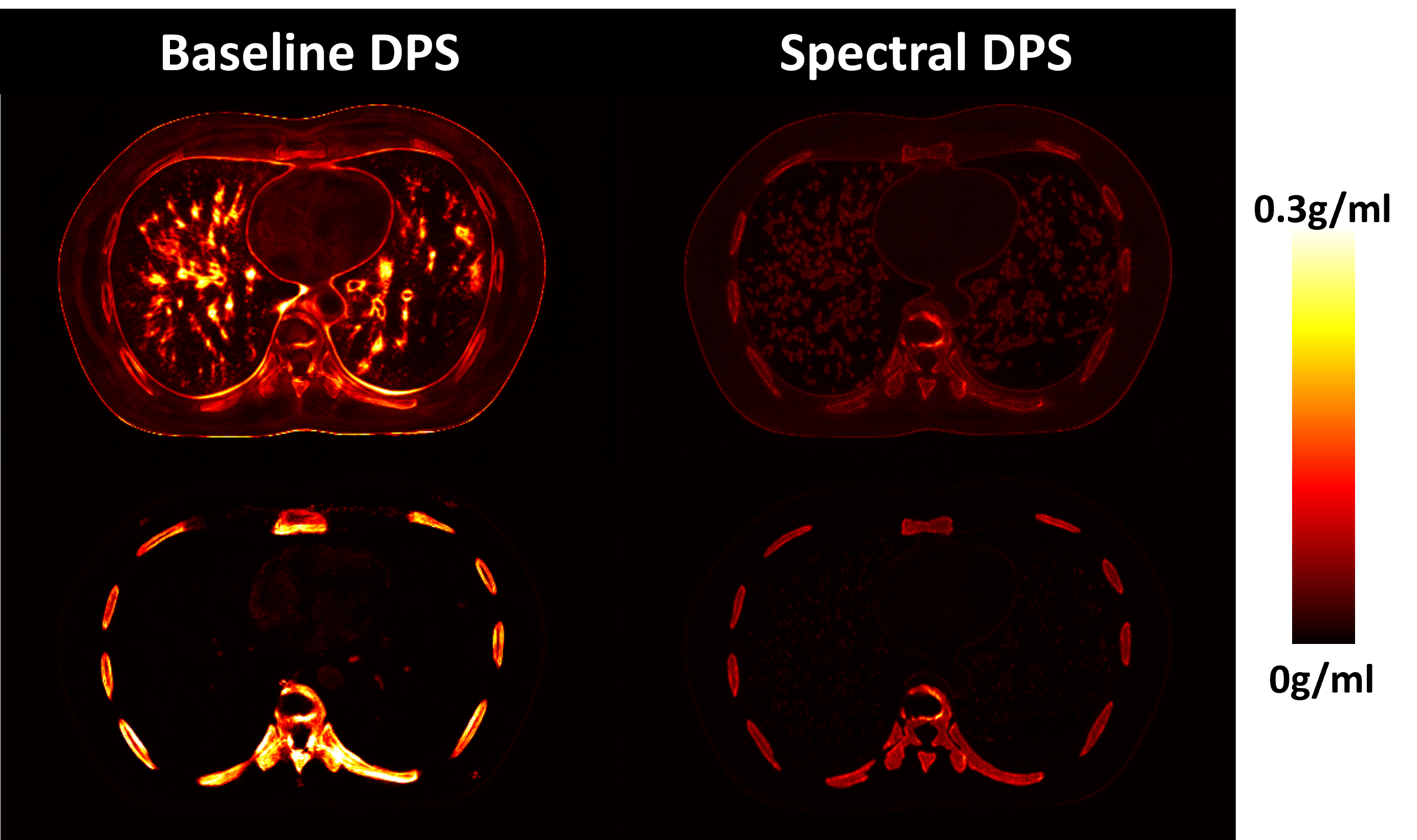}
\centering
\caption{STD map of baseline DPS and spectral DPS evaluated on 32 posterior samplings from bench data. Top and bottom row represents water and calcium, respectively.}
\label{fig:real_std} 
\end{figure}

Decomposition results for the anthropomorphic lung phantom are summarized in Fig.\ref{fig:real}. Note that the ground truth is not available for this case. Instead we provide a single-energy FBP image and note that the only calcium in this phantom is in the bone regions. We do not add any regularization to MBMD to approximate an unbiased maximum likelihood estimator. A homogeneous area marked with a red box was used to quantify errors. Using single-energy FBP image as a reference, it is evident that spectral DPS effectively preserved most of the lung branches in the water image and accurately delineated the boundary of the cortical bone in the calcium image. Both IDD and MBMD exhibited extensive decomposition noise. In the homogeneous heart region, the difference of mean value among MBMD, baseline DPS, and spectral DPS was less than $1\%$, highlighting only minor discrepancies in mean estimation. Notably, spectral DPS achieved a significant noise reduction of $52.35\%$ as compared to baseline DPS, and an even greater reduction when compared to MBMD. Additionally, we note that baseline DPS had a tendency to introduce false structures in the homogeneous phantom. In particular, the heart boundaries and water/fat boundaries in the chest wall. Spectral DPS more reliably visualized those homogeneous regions. The noise advantages were also observed in the calcium images. The STD maps of posterior sampling are presented in Fig.\ref{fig:real_std}. Baseline DPS exhibited considerable uncertainty in capturing finer details, as evidenced by the bright areas in the rib cage and lungs. In contrast, spectral DPS variations were much smaller around the edges, substantially enhancing the stability and reliability of the decomposition of smaller objects.

\section{Discussion and Conclusion}
\label{sec:dis}
In this work, we propose a novel spectral diffusion posterior sampling framework for multi-material decomposition. One of the major advantages of DPS is to decouple network training from specific system configurations, allowing a one-time training for different applications. This flexibility has been validated through the simulation study on the kV-switching and the dual-layer system which have considerably dissimilar spectral sensitivities. Moreover, we extended our evaluation of spectral DPS to a physical system and an anthropomorphic phantom, whose partially realistic features do not perfectly align with the trained prior distribution. Unlike baseline DPS, which is prone to generate non-existent adipose/muscle features, spectral DPS successfully depicts the homogeneous soft tissue in the real phantom. This behavior demonstrates that spectral DPS can maintain a reasonable balance between prior knowledge and measurement information, which enhances its capability of accommodating out-of-distribution samples. Such adaptability further validates the robustness of spectral DPS as well as its potential for broader application in various clinical settings.

Spectral DPS incorporated several strategies from our previous work to further stabilize and accelerate the decomposition. The proposed spectral DPS has demonstrated significantly improved stability and imaging accuracy compared to alternate approaches across both simulated dual-layer and kV-switching systems, as well as physical bench CBCT data. The modified strategies borrow significantly from prior model-based algorithm developments. For example, model-based material decomposition often necessitates extensive iterations to achieve convergence, primarily due to its inherent ill-conditioned nature. For the same reason, baseline DPS with a naive gradient descent optimizer struggles to effectively incorporate physics information into the decomposition process, resulting in significant variability in the outcomes. The jumpstart sampling, initiated from image-domain decomposition, provides an excellent initialization for the reverse process and significantly shortens the number of sampling steps and reduces the potential for divergence caused by the stochastic sampling processes. Additionally, the integration of the Adam optimizer with ordered subset strategy effectively enhances data consistency. While these improvements are substantial, exploring other sophisticated classic optimizers, such as SPS\cite{tilley2019model} and the preconditioned algorithm\cite{tivnan2020preconditioned}, may potentially accelerate convergence even further. The impact of such methods on spectral DPS is left for future investigation. 

One major observation in this work was that a careful parameter optimization seeking minimized STD can not only significantly reduce DPS variability, but that minimum STD is correlated with improved performance on other metrics quantifying the decomposition accuracy, e.g., PSNR and SSIM. Also, the optimal parameters appeared to be similar for different anatomical slices. This consistency may obviate the need for fine tuning, streamlining the decomposition process and enhancing overall efficiency. Future efforts will investigate if such optimal parameters extend to more diverse anatomical targets beyond thoracic CT.

Some clinical applications require further material separation for precise quantification of water, calcium, and exogenous contrast agent, necessitating the decomposition of three or more basis materials. While our current work primarily explores two-material decomposition, extending spectral DPS to more materials should be a straightforward application of the methods developed here. For example, photon-counting CT can potentially provide additional spectral channels (i.e., energy bins) to permit estimation of more material bases. Application to such systems represents future work that is enabled by the proposed spectral DPS approach. We expect that the strategies proposed here can be further modified and applied to related measurement models and other medical imaging modalities.

% use section* for acknowledgment
\section*{Acknowledgments}
This work is supported, in part, by NIH grant R01EB030494.

\bibliography{report} % bibliography data in report.bib
\bibliographystyle{IEEEref}

\end{document}